\documentclass[final,natbib,runninghead]{svjour2}
\usepackage{graphicx}
\newcommand{\bm}[1]{{\mbox{\boldmath $#1$}}} 
\journalname{Earth, Moon, and Planets}
\bibpunct{[}{]}{;}{n}{}{,} 
\begin{document}
\title{Geo-neutrinos: A systematic approach to uncertainties and correlations}
\titlerunning{Geo-neutrinos: A systematic approach}        
\author{	G.L.~Fogli   \and
       	E.~Lisi		\and
		A.~Palazzo \and
		A.M.~Rotunno} 


\institute{	G.L.~Fogli$^1$, 
			E.~Lisi$^1$ (speaker), 
			A.~Palazzo$^{1,2}$, 
			A.M.~Rotunno$^{1}$ 
			\at
            $^1$Dip.\ di Fisica and Sezione INFN di Bari, Via Amendola 173, 70126, Bari, Italy \\
			$^2$	Astrophysics, Denys Wilkinson Building, Keble Road, OX13RH, Oxford, UK}

\date{Received: ..........}

\maketitle

\begin{abstract} Geo-neutrinos emitted by heat-producing elements (U, Th and K)
represent a unique probe of the Earth interior. 
The characterization of their fluxes is subject, however, to 
rather large and highly correlated uncertainties. The geochemical
covariance of the U, Th and K abundances in various Earth reservoirs induces positive correlations
among the associated geo-neutrino fluxes, and between these and the radiogenic heat.
Mass-balance constraints in the Bulk Silicate Earth (BSE) tend instead to
anti-correlate the radiogenic element abundances in complementary reservoirs. 
Experimental geo-neutrino observables may be further (anti)correlated by instrumental effects.
In this context, we propose a systematic 
approach to covariance matrices, based on the fact that all the relevant geo-neutrino 
observables and constraints can be expressed as linear functions of the  U, Th and K 
abundances in the Earth's reservoirs (with relatively well-known coefficients). 
We briefly discuss here the construction of a tentative ``geo-neutrino source model'' (GNSM) 
for the U, Th, and K abundances in the main Earth reservoirs, 
based on selected geophysical and geochemical data and models 
(when available), on plausible hypotheses (when possible), and admittedly on arbitrary 
assumptions (when unavoidable). We use then the GNSM to make predictions about
several experiments (``forward approach''), and to show how future data
can constrain {\em a posteriori\/} the error matrix of the model itself (``backward approach'').
The method may provide a useful statistical framework for evaluating the impact 
and the global consistency of prospective geo-neutrino measurements and Earth models.

\end{abstract}

\section{Introduction}
\label{sec:intro}

Electron antineutrinos emitted in the decay chains of the heat-producing 
elements (HPE) U, Th, and K in terrestrial rocks --- the so-called geo-neutrinos ---
represent a truly unique probe of the Earth interior; see \cite{Fi05a} for a 
recent review and \cite{Kr84} for earlier discussions and references. The first indications
for a (U+Th) geo-neutrino signal at $>2\sigma$ confidence level in the KamLAND experiment by
\cite{Ar05}
have boosted the interest in this field, and have started to bridge the two 
communities of particle physicists and Earth scientists --- as exemplarily testified  
by this Workshop (Neutrino Geophysics, Honolulu, Hawaii, 2005). 

The hope is that future measurement of geo-neutrino fluxes can put statistically 
significant constraints to the global abundances of HPE's and to their associated heat 
production rates, which are currently subject to highly debated
Earth model assumptions \cite{Mc03,Sleep}. This goal, despite being experimentally
very challenging, is extremely important and deserves dedicated (possibly
joint) studies from both scientific communities.

One methodological difficulty is represented by the different ``feeling'' for uncertainties
by particle physicists {\em vs\/} Earth scientists. First important attempts to systematize
(U, Th, K) abundance uncertainties in a format convenient for geo-neutrino analyses
have been performed in \cite{En05} and particularly in \cite{Ma05,Ma04,Fi05b}, 
where errors have been basically assessed
 from the spread in published estimates (consistently with mass balance
constraints). 

We propose to make a further step, by systematically taking into account the
ubiquitous error {\em covariances}, i.e., the fact that several quantities 
happen to vary in the same direction (positive correlations) or in opposite directions
(negative correlations) in the geo-neutrino context.  For instance, in a given Earth
reservoir (say, the mantle), the U, Th and K abundances are typically positively correlated.
However, they may be anticorrelated in two complementary reservoirs constrained 
by mass balance arguments, such as the mantle and the crust in Bulk Silicate Earth
(BSE) models. 
Experimental geo-neutrino observables may be further (anti)correlated by instrumental effects.

An extensive discussion of our approach to these problems is beyond the scope
of this contribution and will be presented elsewhere \cite{Ours}. Here we briefly
report about some selected issues and results, according to the following scheme. 
In Sec.~2 we discuss the general aspects and the statistical tools related
to covariance analyses, with emphasis on geo-neutrino observables. In Sec.~3 we
construct a tentative model for the source distribution of (U, Th, K) in global
Earth reservoirs (Geo Neutrino Source Model, GNSM). In Sec.~4 we discuss some issues
related to the characterization of local sources around geo-neutrino detector sites.
In Sec.~5 and 6 we show examples of ``forward'' error estimates 
(i.e., propagation of GNSM errors to predicted geo-neutrino rates) 
and shortly discuss ``backward'' error updates
(i.e., GNSM error reduction through prospective geo-neutrino data).
We draw our conclusions in Sec.~7.

\section{Covariance and correlations: General aspects}
\label{sec:cova}

In this Section we discuss some general aspects of covariance analyses in geochemistry and
in neutrino physics, and then present the basic tools relevant for geo-neutrino physics.
We remind that, for any two quantities $P$ and $Q$, estimated as 
\begin{eqnarray}
P&=&\overline P \pm\sigma_P\ , \\ 
Q&=&\overline Q\pm\sigma_Q\ ,
\end{eqnarray}
the correlation index $\rho_{PQ}\in[-1,+1]$ 
between the 1-$\sigma$ errors of $P$ and $Q$ parameterizes
the degree of ``covariation'' of the two quantities: $\rho>0$ ($<0$) if they change in
the same (opposite) direction, while $\rho=0$ if they change independently;
see, e.g.\ \cite{Stat}. The 
covariance (or squared error) matrix of $P$ and $Q$ contains $\sigma^2_P$ and $\sigma^2_Q$
as diagonal elements, and $\rho_{PQ}\sigma_P\sigma_Q$ as off-diagonal ones.
For more than two variables with errors $\sigma_i$ and correlations
$\sigma_{ij}$, the covariance matrix is $\sigma^2_{ij}=\rho_{ij}\sigma_i\sigma_j$
(symmetric, with $\rho_{ii}=1$ on the diagonal).

\subsection{Covariance analyses in geochemistry}

In 1998, the Geochemical Earth Reference Model (GERM) initiative was officially launched
\cite{GERM}, in order to provide a ``consensus model'' for the elemental abundances, together
with their errors
and correlations, in all relevant Earth reservoirs. Although a lot of work has been done
in this direction, e.g., through rich compilations of data and estimates 
(www.earthref.org), the correlation 
matrices have not yet been estimated --- not even for subsets of elements such as (U, Th, K).
To our knowledge, only a few regional studies discuss HPE covariances. 
These difficulties can be in part overcome by using the (more frequently reported)
elemental ratio information. For instance, if the ratio of two abundances $P$ and $Q$
is reported together with its error $\sigma_{P/Q}$, the correlation between $P$ and $Q$
can be inferred through the following
statistical relation, valid at first order in error propagation:
\begin{equation}
\label{ratio}\left(\frac{\sigma_{P/Q}}{P/Q}\right)^2=\Big(\frac{\sigma_{P}}{P}\Big)^2
+\Big(\frac{\sigma_{Q}}{Q}\Big)^2
-2\rho_{PQ}\Big(\frac{\sigma_{P}}{P}\Big)
\Big(\frac{\sigma_{Q}}{Q}\Big)\ .
\end{equation}
Although the ``ratio'' and ``correlation'' information appear to be interchangeable through
the above formula, from a methodological viewpoint it is better to use the latter rather
than the first,
since the ratio of two Gaussian variables is 
a Cauchy distribution with formally
infinite variance \cite{Stat} --- a rather tricky object in statistical manipulations.

\subsection{Covariance analyses in neutrino physics}

Neutrino physics has undergone a revolution in recent years, after the discovery of neutrino
flavor oscillations. Phenomenological fits to neutrino oscillation data have become
increasingly refined, and now routinely include covariance analyses (see, e.g., 
\cite{PPNP}). 
For our purposes, a relevant example is also given by the Standard Solar Model \cite{Bahc}, which 
provides, among other things, errors and correlations for solar neutrino sources.
We shall try to apply a similar ``format'' to a geo-neutrino source model in Sec.~3.

\begin{figure*}
\centering
  \includegraphics[width=0.5\textwidth]{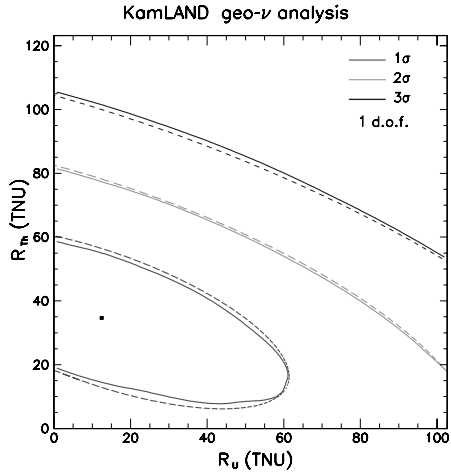}
\caption{Best-fit U and Th event rates and error contours (solid lines) 
from our analysis of KamLAND geo-neutrino data \protect\cite{Ours}. The contours are very close
to two-dimensional Gaussian confidence levels (dashed ellipses). Units: 1 TNU = 
1 event/year/$10^{32}$ target protons. }
\label{fig:1}       
\end{figure*}

Correlations arise not only at the level of neutrino sources, but also at the
detection level, as a consequence of
instrumental effects. For instance, the KamLAND experiment is currently more sensitive
to the sum (U+Th) of geo-neutrino fluxes rather than to the separate U and Th components.
As a consequence, the measured U and Th geo-neutrino event rates are anticorrelated:
if one rate increases the other one tends to decrease, in order to keep the total rate constant 
(within errors).

Figure~1 shows explicitly the anticorrelation
between the U and Th experimental rates through 
their 1, 2, and 3-$\sigma$ contours (solid lines) taken from our KamLAND
 data analysis \cite{Ours}. 
The contours in Fig.~1 are well approximated by a bivariate gaussian (dashed lines)
with parameters:
\begin{equation}
R_\mathrm{U}=12.5\pm 48.9\mathrm{\ TNU},\ R_\mathrm{Th}=34.7\pm 28.5 \mathrm{\ TNU},\
\rho = -0.645\ .  
\end{equation} 
We have also verified that our KamLAND data analysis reproduces the confidence
level contours in the alternative plane spanned by $R_\mathrm{U}+R_\mathrm{Th}$
and $(R_\mathrm{U}-R_\mathrm{Th})/(R_\mathrm{U}+R_\mathrm{TH})$ \cite{Ar05} (not shown); 
however, as explained
in the previous subsection, we prefer to avoid
any ``ratio'' and to use just $(R_\mathrm{U},\,R_\mathrm{Th})$ and their correlation.

\subsection{General tools for geo-neutrinos}

In the context of geo-neutrinos, statistical analyses are greatly simplified by the fact 
that all relevant observables are linear combinations of the HPE abundances $a^\mathrm{S}_i$  in
different reservoirs (S~=~U, Th, K; $i=$ reservoir index), 
with coefficient determined by known physics and by the geometry of the Earth
mass distribution.

\begin{table}[t]
\caption{\label{Univ}  Universal (reservoir-independent and detector site-independent) coefficients for the calculation of 
the total  heat ($H$)  of the Earth, and of the total $\overline\nu_e$ flux 
($\Phi_D$) and event rates from inverse beta decay ($R_D$) at any detector site $D$. 
Conversion factors: 1 TNU = 1 event/year/$10^{32}$ target protons; 1 year = $3.15576\times 10^{7}$ s. 
Natural abundances of isotopes are assumed.}
\centering
\begin{tabular}{cccc}
\hline\noalign{\smallskip}
			&	$h_\mathrm{S}$		&	$\phi_\mathrm{S}$				&	$r_\mathrm{S}$				\\
S			& ($\mu$W/kg)			& ($10^{12}$ $\overline\nu_e$/cm$^2$/s)	& ($10^8$ TNU)				\\[1mm]
\tableheadseprule\noalign{\smallskip}
U			& 98.0					& 	123								& 15.2							\\
Th			& 26.3					&	26.1								&	1.06							\\					
K			&$34.9\times 10^{-4}$	&	$45.4\times10^{-3}$ 				&    0			 				 \\	
\noalign{\smallskip}\hline
\end{tabular}
\end{table}

In particular we consider: 
($i$) the total radiogenic heat $H$ 
of the Earth (decay energy absorbed per unit of time); 
($ii$) the geo-neutrino flux $\Phi_D$ at a given 
detector site $D$ (number of $\overline\nu_e$ per unit of area and time); and
($iii$) the corresponding event rate
$R_D$ at a given detector site $D$ (number of events from $\overline\nu_e+p\to n+e^+$,
per unit of time and of target protons). Such quantities can be
written as: 
\begin{eqnarray}
\label{H}
H_R 			&=& \sum_\mathrm{S} h_\mathrm{S} \sum_i M_i\, a_i^\mathrm{S}\ ,\\
\label{PhiD}
\Phi_D 		&=& \langle P_{ee}\rangle \sum_\mathrm{S} \phi_\mathrm{S} \sum_i f_i^D\, a_i^\mathrm{S}\ ,\\
\label{RD}
R_D 		&=& \langle P_{ee}\rangle \sum_\mathrm{S} r_\mathrm{S} \sum_i f_i^D\, a_i^\mathrm{S}\ ,
\end{eqnarray}
where the universal coefficients $h_\mathrm{S}$,  $\phi_\mathrm{S}$,
and $r_\mathrm{S}$, according to our calculations \cite{Ours}, are given in Table~\ref{Univ}.
In the above equations, $M_i$ is the mass of the
$i$-th reservoir, while $\langle P_{ee}\rangle\simeq 0.57$ is the average
survival oscillation probability of geo-$\overline\nu_e$. The geometrical coefficients
$f_i^\mathrm{D}$ represent the mass-weighted average of the inverse square distance
of the detector site $D$ from the $i$-th reservoir, necessary to account for the
flux decrease with distance; their numerical values will be reported elsewhere \cite{Ours}.

The Earth mass distribution necessary to compute the $M_i$'s and $f_i^\mathrm{D}$'s 
is taken from the Preliminary
Earth Reference Model (PREM) \cite{PREM}, properly matched with a crustal model 
defined over a grid of $2^\circ\times2^\circ$ tiles \cite{2by2}. The Earth is 
assumed to be partitioned into the following homogeneous reservoirs:
core, lower mantle, upper mantle, continental
crust (in three layers---upper, middle, lower) and oceanic crust (lumped into
one layer). The distinction (if any) between lower and upper mantle is 
strongly debated and will be commented later. For a set of possible
geo-neutrino detector sites (Kamioka, Gran Sasso, Sudbury, Hawaii, Pyh\"asalmi, Baksan),
we consider ``local'' reservoirs, defined as the nine (three-by-three) tiles of
the $2^\circ\times 2^\circ$ model which surround each detector site---except for Kamioka,
where 13 tiles are considered, corresponding to the ``Japanese arc and forearc'' as defined
in the crustal model \cite{2by2}. Due to the inverse-squared-distance decrease of the neutrino
flux, it turns out that local and global reservoirs can provide comparable contributions
to the geo-neutrino event rates, at least for detectors sitting on the continental crust.

The main task is then to build a model for the abundances $a_i^\mathrm{S}$, embedding
covariances. In other words, by switching to a single-index
vector notation for simplicity, 
\begin{equation}
\label{vector} \left\{a^\mathrm{S}_i\right\}^{\mathrm{S=U,Th,K}}_{i=1,\dots,N}\to
\mathbf{a}=\left\{a_i\right\}_{i=1,\dots,3N}\ 
\end{equation}
(where $N$ is the number of reservoirs), an Earth model should 
provide, for any entry in the HPE abundance vector $\mathbf{a}$,
both a central value $\overline a_i$ and a standard deviation $\pm\sigma_i$,
\begin{equation}
\label{errordef} a_i=\overline a_i\pm \sigma_i
\end{equation}
together with the error correlation matrix $\bm{\rho}$. The components of
the covariance matrix 
$\bm{\sigma}^2$ are then  
\begin{equation}
\label{sigmadef}[\bm{\sigma}^2]_{ij}=\rho_{ij}\sigma_i\sigma_j\ .
\end{equation}

Given any two quantities $P$ and $Q$ [as in Eqs.~(\ref{H})--(\ref{RD})]  
defined as linear combinations of the $a_i$'s (with $T=$~transpose),
\begin{eqnarray}
\label{PQ1}P&=&\sum_i p_i a_i=\bm{p}^T \bm{a} \ ,\\
\label{PQ2}Q&=&\sum_i q_i a_i=\bm{q}^T \bm{a}\ ,
\end{eqnarray}
it turns out that their 1-$\sigma$ errors $\sigma_P$ and $\sigma_Q$ are simply given by
\begin{eqnarray}
\label{errPQ1} \sigma^2_P &=& \sum_{ij} p_i p_j \rho_{ij} \sigma_i \sigma_j
=\bm{p}^T\bm{\sigma}^2 \bm{p}\ ,\\
\label{errPQ2} \sigma^2_Q &=& \sum_{ij} q_i q_j \rho_{ij} \sigma_i \sigma_j
=\bm{q}^T\bm{\sigma}^2 \bm{q}\ ,
\end{eqnarray}
with $(P,Q)$ correlation given by
\begin{equation}
\label{rhoPQ} \rho_{PQ}=\frac{\sum_{ij}p_iq_j\rho_{ij}\sigma_i\sigma_j}{\sigma_P\sigma_Q}
=\frac{\bm{p}^T\bm{\sigma}^2 \bm{q}}{\sigma_P\sigma_Q}\ .
\end{equation}
The above equations will be used below to compute correlations among 
experimental event rates, or between an experimental rate and the radiogenic heat.

\section{Towards a Geo-Neutrino Source Model (GNSM)}
\label{sec:gnsm}

In this Section we briefly discuss our methodology to provide entries for 
Eqs.~(\ref{errordef},\ref{sigmadef}), i.e., a Geo-Neutrino Source Model
for HPE abundances in the Earth. We remind that, concerning the entries
for Eq.~(\ref{errordef}) 
(errors only, no correlations), our approach overlaps in part with earlier relevant work 
performed in \cite{Fi05a,Ma05,Fi05b}.

\subsection{The Bulk Silicate Earth}
Bulk Silicate Earth (BSE) models \cite{Mc03} provide global constraints on elemental
abundances (especially in the primitive mantle), under a set of hypotheses. 
In particular, BSE models include the plausible
assumption that elements which have both high condensation temperature 
(``refractory'') and that are preferentially embedded in rocks rather in iron (``lithophile'') 
should be found in the primitive mantle (i.e., in the undifferentiated mantle+crust reservoir) 
in the same ratio as in the parent, pristine meteoritic material. 
Among the three main HPE's (U, Th, K), the first two are also refractory lithophile
elements (RLE), so the Th/U global ratio should be the same in BSE and in the 
(supposedly) parent and most primitive meteoritic material (carbonaceous chondrites CI).
Our summary \cite{Ours} of the recent and detailed works on absolute \cite{Palm,Lodd} and relative
\cite{Roch,Gore} U and Th abundances in CI meteorites (with $1\sigma$ errors) is: 
\begin{eqnarray}
a^\mathrm{Th}_\mathrm{CI}&=&30.4(1\pm 0.10) \times 10^{-9}\ ,\\
a^\mathrm{U}_\mathrm{CI}&=&8.10(1\pm 0.10)\times 10^{-9}\ ,\\
\left(a^\mathrm{Th}_\mathrm{CI}/a^\mathrm{U}_\mathrm{CI}\right)&=&3.75(1\pm 0.05)\ ,
\end{eqnarray}
which implies a (Th,U) error correlation $\rho_\mathrm{CI}=0.875$ through Eq.~(\ref{rhoPQ}).

The BSE/CI abundance ratio is expected to be the same for all RLE's, if indeed
they did not volatilize during the Earth formation history. The benchmark is usually provided
by a major RLE element such as Al, which,
being much more abundant than the trace elements Th and U, can be more robustly
constrained, both by mass-balance arguments and by direct sampling. 
Our summary \cite{Ours} for the BSE/CI abundance ratio of Al from three detailed
BSE models \cite{Palm,Alle,MSun} is 
\begin{equation}
\left(a^\mathrm{Al}_\mathrm{BSE}/a^\mathrm{Al}_\mathrm{CI}\right)=2.7(1\pm 0.10)\ .
\end{equation}
The previous arguments and estimates imply that
\begin{eqnarray}
a^\mathrm{Th}_\mathrm{BSE}=a^\mathrm{Th}_\mathrm{CI}\cdot
\left(a^\mathrm{Al}_\mathrm{BSE}/a^\mathrm{Al}_\mathrm{CI}\right)
&=&82.1(1\pm 0.14) \times 10^{-9}\ ,\\
a^\mathrm{U}_\mathrm{BSE}=a^\mathrm{Th}_\mathrm{CI}\cdot
\left(a^\mathrm{Al}_\mathrm{BSE}/a^\mathrm{Al}_\mathrm{CI}\right)
&=&21.9(1\pm 0.14)\times 10^{-9}\ ,
\end{eqnarray}
with (Th,U) error correlation $\rho_\mathrm{BSE}=0.936$.

The K element, being (moderately) volatile, needs a separate discussion. In
\cite{Joch} it was argued that U is a good ``global'' proxy for K, since: (1) the K/U
abundance ratio was found to be nearly constant in 22 samples of Mid-Ocean Ridge Basalt
(MORB) from the Atlantic and Pacific ocean floor (thought to be
representative of the whole mantle); (2) the MORB K/U ratio was
found to be (accidentally) similar to the K/U ratio estimate in the crust from an
older model \cite{Wass}.  The MORB K/U ratio ($1.27\times 10^4$) was then boldly
generalized to the whole Earth, with small estimated errors (1.6\%) \cite{Joch}. 
However, it should be noticed there are no geochemical arguments to
presume that disparate elements such as U and K should have the
same partition coefficients between melt (=~crust) and residual mineral 
(=~depleted mantle); indeed, analogous alleged coincidences have later been
disproved \cite{Hofm}. Therefore, we think that
the ``canonical K/U=12,700'' ratio, so often quoted
in the geochemical literature, should be critically revisited in future studies. Provisionally,
from a survey of recent literature about the abundances
of K and U (and of another possible K-proxy element, La \cite{Palm}) in
MORB databases, continental crust samples and estimates, and
BSE models, we are inclined to \cite{Ours}: (1) increase
significantly --- although subjectively --- the K/U uncertainty; and (2) slightly lower the
central value (as compared with \cite{Joch}). More precisely, we take 
\begin{equation}
\left(a^\mathrm{K}_\mathrm{BSE}/a^\mathrm{U}_\mathrm{BSE}\right)=1.2\times 10^4(1\pm 0.15)\ ,
\end{equation}
which, by proper error propagation, gives the absolute K abundance as
\begin{equation}
a^\mathrm{K}_\mathrm{BSE}=263\times 10^{-6}(1\pm 0.21)\ ,
\end{equation}
with (K,Th) and (K,U) correlations equal to 0.648 and 0.701, respectively.
Table~2 presents a summary of the BSE (U, Th, K) abundances, errors and correlation matrix, 
together with similar information about the main BSE sub-reservoirs (as discussed below).

Needless to say, all the above BSE estimates may be significantly altered, 
if possible indications for nonzero HPE abundances in the Earth core \cite{Rama}
are corroborated by further studies. For the sake of simplicity,
we do not consider such possibility in this work.

\begin{table}[t]
\caption{\label{Corr} Geo-Neutrino Source Model (GNSM):  Abundances, errors and correlations of radiogenic elements
(U, Th, K) in global reservoirs. }
\centering
\resizebox{\textwidth}{!}{
\begin{tabular}{lclc|ccc|ccc|ccc|ccc|ccc}
\hline\noalign{\smallskip}
\multicolumn{4}{c|}{Geo-Neutrino Source Model (GNSM)}	& \multicolumn{15}{c}{Correlation matrix} \\
\multicolumn{4}{c|}{ for global reservoirs }     	    	&\multicolumn{3}{c|}{BSE}&\multicolumn{3}{c|}{CC}&\multicolumn{3}{c|}{OC}&\multicolumn{3}{c|}{UM}&\multicolumn{3}{c}{LM}\\ [1mm]
\tableheadseprule\noalign{\smallskip}
Reser. 	& Elem. & Abundance  		&$\pm1\sigma$ 	& U     & Th     & K    		& U      & Th     & K    	& U	     & Th     & K    	& U      & Th     & K    	& U      & Th     & K    	\\[1mm]
\hline
       	& U 		& $21.9\times 10^{-9}$ 	&$\pm14\%$	&~~~1~~~&$+.936$ &$+.701$ 	&0&0&0 						&0&0&0 						&0&0&0 						&$+.908$ &$+.893$ &$+.690$ 	\\
  BSE   	& Th 	& $82.1\times 10^{-9}$ 	&$\pm14\%$ 	&       &   1    &$+.648$ 	&0&0&0 						&0&0&0 						&0&0&0 						&$+.850$ &$+.954$ &$+.638$ 	\\
	 	& K 		& $26.3\times 10^{-5}$ 	&$\pm21\%$ 	&       &        &   1 		&0&0&0 						&0&0&0 						&0&0&0 						&$+.636$ &$+.618$ &$+.985$ 	\\[1mm]
\hline
       	& U 		& $1.46\times 10^{-6}$ 	&$\pm17\%$	&&& 					 		&~~~1~~~ &$+.906$ &$+.906$  	&0&0&0 						&0&0&0 						&$-.409$ &$-.263$ &$-.146$ 	\\
  CC   	& Th 	& $6.29\times 10^{-6}$ 	&$\pm10\%$ 	&&& 							&        &   1    &$+.595$ 	&0&0&0 						&0&0&0 						&$-.371$ &$-.291$ &$-.096$ 	\\
	 	& K 		& $1.62\times 10^{-2}$ 	&$\pm10\%$ 	&&& 							&        &        &   1    	&0&0&0 						&0&0&0 						&$-.371$ &$-.173$ &$-.161$ 	\\[1mm]
\hline
       	& U 		& $1.00\times 10^{-7}$ 	&$\pm30\%$	&&& 							&&&							&~~~1~~~ &$+.906$ &$+.868$ 	&0&0&0 						&$-.012$ &$-.007$ &$-.007$ 	\\
  OC   	& Th 	& $2.20\times 10^{-7}$ 	&$\pm30\%$ 	&&& 							&&& 							&        &  1     &$+.764$ 	&0&0&0 						&$-.011$ &$-.007$ &$-.006$ 	\\
	 	& K 		& $1.25\times 10^{-3}$ 	&$\pm28\%$ 	&&&							&&& 							&        &        &   1   	&0&0&0 						&$-.010$ &$-.006$ &$-.008$ 	\\[1mm]
\hline
       	& U 		& $3.95\times 10^{-9}$ 	&$\pm30\%$	&&& 							&&& 							&&& 							&~~~1~~~ &$+.906$ &$+.868$ 	&$-.093$ &$-.065$ &$-.058$ 	\\
  UM   	& Th 	& $10.8\times 10^{-9}$ 	&$\pm30\%$ 	&&& 							&&& 							&&& 							&        &   1    &$+.764$ 	&$-.084$ &$-.071$ &$-.051$ 	\\
	 	& K 		& $5.02\times 10^{-5}$ 	&$\pm28\%$ 	&&& 							&&& 							&&&							&        &        &   1   	&$-.081$ &$-.054$ &$-.066$ 	\\[1mm]
\hline
       	& U 		& $17.3\times 10^{-9}$ 	&$\pm27\%$	&&& 							&&& 							&&&	 						&&& 							&~~~1~~~ &$+.924$ &$+.692$ 	\\
  LM   	& Th 	& $60.4\times 10^{-9}$ 	&$\pm27\%$ 	&&& 							&&& 							&&& 							&&& 							&        &   1    &$+.640$ 	\\
	 	& K 		& $21.7\times 10^{-5}$ 	&$\pm36\%$ 	&&& 							&&& 							&&& 							&&& 							&        &        &   1 \\[1mm]
\noalign{\smallskip}\hline
\end{tabular}
}
\end{table}

\subsection{The Continental Crust (CC)}

Average elemental abundances in the continental crust (CC), and their vertical 
distribution in the three main identifiable layers (upper, middle, lower crust = UC, MC, LC), 
have been presented in a recent comprehensive review \cite{Rudn}, together
with a wealth of data
and with a critical survey of earlier literature on the subject.  In 
particular, it is stressed in \cite{Rudn} that some previous CC models are not consistent 
with known crustal heat production constraints \cite{Jaup}. This fact shows that: (1) the
spread of published values for elemental abundances
is not necessarily indicative of the real uncertainties, since 
some estimates can be invalidated by new and independent
data; (2) heat production estimates in the CC provide a relevant constraint  
(linear in the U, Th, K crustal abundances) which might help, together
with geo-neutrino measurements, to reduce the HPE abundance estimates in reference models.
The latter point will be further elaborated elsewhere \cite{Ours}.

We basically adopt the results in \cite{Rudn} for the UC, MC, LC abundances of (U,~Th,~K)
and their uncertainties, with the following differences: (1) since no error estimates
are given for the LC, we conservatively (but arbitrarily)
assume fractional $1\sigma$ errors of $40\%$
in this layer; (2) our reference crustal model \cite{2by2} and the one in \cite{Rudn}
provide mass ratios among layers 
(UC:MC:LC) respectively equal to 0.359:0.330:0.311 and 0.317:0.296:0.387. This
difference is somewhat disappointing, since it induces weighted-average
HPE abundance shifts in the CC of order $10\%$, which are definitely nonnegligible. 
``Consensus values'' for the mass distribution in the three CC layers (upper, middle, lower)
would thus be desirable in the future. Provisionally, we assume that CC
elemental abundance errors cannot be smaller than the 
``mass distribution'' induced error (10\%). We also assume, from a 
survey of the relevant literature, a $9\%$ fractional error for each of the
K/U, Th/U, and K/Th ratios in the crust --- which in turn provide the (U, Th, K) correlations
\cite{Ours}.  Given such inputs, the CC abundances (central values, errors, and correlations) 
turn out to be as shown in Table~2.

\subsection{The Upper Mantle (UM)}

We assume a homogeneous composition for the upper mantle (defined as the sum
of transition zone + low-velocity zone + ``lid'' in the PREM model \cite{PREM}). Global and detailed
analyses of all the available upper mantle samples and constraints have been performed
in two recent papers \cite{Salt,Work} which, unfortunately, do not really agree in their conclusions,
despite being based in part on the same petrological database (www.petdb.org).
Concerning HPE's, we then take as central values the average of \cite{Salt} and \cite{Work}, but
we attach the most conservative error estimates of \cite{Salt}, which are large enough to cover
the spread between \cite{Salt} and \cite{Work}. 
We assume a K/U ratio error in the upper mantle of the
same size as for the BSE ($\pm15\%$), and a Th/U ratio error of $\pm13\%$, as suggested
from the scatter of points in Fig.~2 of \cite{Salt}. Given such inputs, the UM abundances
turn out to be \cite{Ours} as shown in Table~2.

\subsection{The Oceanic Crust (OC)}

The oceanic crust is difficult to sample and, not surprisingly, only a few papers
(to our knowledge) deal with its average trace-element composition  (see, e.g., \cite{OCTa,OCHo,OCWe}). We adopt the 
``intermediate'' central values of \cite{OCTa}, which suggest a HPE enrichment of the 
oceanic crust by a factor 20--25 with respect to the (parent) upper mantle. Since the
enrichment is approximately uniform for all three HPE's, we think it reasonable 
to assume that the same relative spread of abundances is transferred from the  
UM (parent mineral) to the OC (melt). Therefore, in the absence of other information, 
we attach to the (U, Th, K) abundances in the
OC the same fractional errors and correlations as for the UM, see Table~2.

\subsection{The Lower mantle (LM)}

The consistent 
derivation of LM abundances, errors and correlations is a qualifying result of
our work. The abundances in the lower mantle (LM) are obtained by subtraction (LM=BSE-UM-CC-OC),
namely, by the mass balance constraint:
\begin{equation}
a^\mathrm{S}_\mathrm{LM}=(a^\mathrm{S}_\mathrm{BSE} M_\mathrm{BSE}
-a^\mathrm{S}_\mathrm{UM} M_\mathrm{UM}-a^\mathrm{S}_\mathrm{CC} M_\mathrm{CC}
-a^\mathrm{S}_\mathrm{OC} M_\mathrm{OC})/M_\mathrm{LM}
\ ,
\end{equation}
for S=U, Th, K. Since the three HPE abundances $a^\mathrm{S}_\mathrm{LM}$ are linear
combinations of BSE, UM, CC, and OC abundances, it is possible to apply the formalism
of Sec.~2.3 and to obtain their errors and correlations, whose numerical values
 are listed in Table~2
(last three columns). It turns out that the LM fractional uncertainties are comparable to those
of the UM ($\sim 30\%$), and that the LM abundances are strongly correlated with
the BSE ones
but moderately anticorrelated with the CC ones, due to the subtraction procedure. 
The LM anticorrelation with the UM and OC is instead very small, since the latter
two reservoirs contain relatively small absolute amounts of HPE's, as compared
to the CC and BSE. 

Once the LM contents of HPE's are obtained, the BSE information becomes redundant,
and one can proceed with the information contained in the 12 entries for the
(CC, OC, UM, LM) abundances, and the corresponding
$12\times12$ correlation matrix, which are reported in Table~2. Notice that,
within the quoted
uncertainties, the abundances in Table~2 are consistent with those reported in \cite{Ma04}.

\subsection{What about Mantle Convection?}

There is currently a strong debate about the nature and extent of mantle convection,
with scenarios ranging from two-layer models (with geochemically decoupled UM and LM) 
to whole mantle convection (with completely mixed UM and LM), and many intermediate
possibilities and variants \cite{Mc03,Roma}. 
Two extreme possibilities are: (1) a geochemically
homogeneous mantle (i.e., no difference between UM and LM, 
$a^\mathrm{S}_\mathrm{LM}=a^\mathrm{S}_\mathrm{UM}$); and
(1) a strict two-layer model (i.e., a lower mantle conserving
primitive mantle abundances,
$a^\mathrm{S}_\mathrm{LM}=a^\mathrm{S}_\mathrm{BSE}$).

Our estimates in Table~2 are intermediate between such two cases, and thus
agree better with models predicting partial mantle mixing. The two extreme cases 
 are anyway
recovered by stretching the uncertainties to roughly $\pm 3\sigma$. Fig.~2 shows
the 1, 2, and $3\sigma$ error ellipses in the (LM,UM) and (LM,BSE) Uranium abundance
planes; within $3\sigma$, both cases $a^\mathrm{U}_\mathrm{LM}=a^\mathrm{U}_\mathrm{UM}$
and  $a^\mathrm{U}_\mathrm{LM}=a^\mathrm{U}_\mathrm{BSE}$ (slanted lines) are allowed. 
Similar results are obtained for Th and K (not shown).
Therefore, our GNSM estimates are sufficiently conservative to cover, within 
$\pm 3\sigma$, a wide spectrum of mantle mixing scenarios
(two-layer convection, partial UM-LM mixing, whole mantle convection).

\begin{figure*}
\vspace*{4mm}
\centering
  \includegraphics[width=0.99\textwidth]{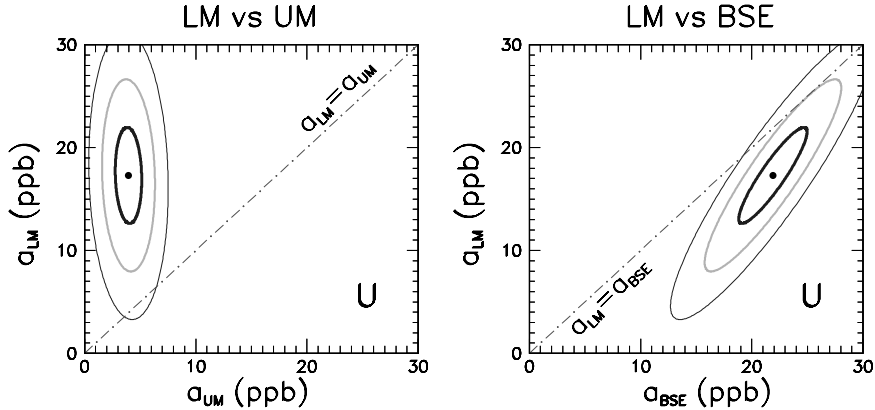}
\vspace*{2mm}
\caption{Comparison between LM, UM, and BSE abundances of Uranium, in ppb ($10^{-6}$) 
units. Our GNSM 
estimates are shown as 1, 2, and 3$\sigma$ error ellipses. The slanted lines
represent the cases of whole mantle convection (in the left panel) and 
of decoupled  LM (in the right panel).}
\label{fig:2}       
\vspace*{3mm}
\end{figure*}

\section{Issues related to ``local'' reservoirs}
\label{sec:local}

In our work, local reservoirs have been arbitrarily defined as the nine crustal 
tiles around each detector (except for Kamioka), see Sec.~2.3. Here we discuss some
issues related to this or other choices for the ``local'' contribution to geo-neutrino
fluxes.

\subsection{What is a ``local'' reservoir?}

It is necessary to define in some way ``local'' reservoirs, since the crust
(and perhaps the mantle) within a few hundred km from each detectors site may well be
different from the average crust defined in the previous section. This fact has
already been recognized in \cite{En05,Fi05b}, where ``local'' HPE abundances for the
Kamioka site have been estimated. The boundaries of the local crust are
matter of convention --- but any convention is not without consequences, however.
In particular, in our approach, we observe that the
correlation between ``local'' reservoirs and ``global'' ones (LM, UM, OC, CC)
is expected to vanish: the uncertainty of the U abundance near the Kamioka mine
has probably nothing to do with the errors of the whole CC and OC crust estimates.
However, only dedicated
studies, which should
take into account all locally homogenous geochemical micro-reservoirs
and their correlation lengths with farther geo-structures, can provide
a physically motivated distinction between local and global reservoirs---a task much beyond 
the scope of this work.
For simplicity,
we just assume that all local-to-global abundance
correlations are exactly zero.

\subsection{Horizontal and vertical (U, Th, K) distribution uncertainties}

Assuming that a ``local'' reservoir is defined in some way, its volumetric
distributions of HPE's can significantly affect the estimated
geo-$\nu$ fluxes, due to the inverse square distance dependence. In principle,
one would like to have such detailed information  around each detector site, both
horizontally and vertically.
In practice, however, one usually has mainly scattered ``surface'' samples and 
only weak constraints about the vertical HPE distribution. Although the 
HPE abundances are expected to decrease with depth, the decrease may
be highly site-dependent and non-monotonic (see \cite{LocJ} as an example for the
Japanese crust). In some cases
(e.g., Sudbury \cite{LocS}) the horizontal and vertical distributions of HPE's
are correlated by the fact that the crust is locally ``tilted'' --- a 
situation which may represent both a complication and an opportunity. 

In all cases, significant progress in the characterization of the HPE
volumetric distribution in local reservoirs can be obtained only by
dedicated geophysical and geochemical studies, which should collect all the (currently sparse
and partly unpublished) relevant pieces of data, including representative
rock samples, local crustal models, and heat flow measurements.
Some interesting work in this direction has been done for the Kamioka site \cite{En05,Fi05b}, 
showing that a $O(10\%)$ uncertainty in the local geo-neutrino flux (at $1\sigma$)
may perhaps be reachable. We think that $10\%$ 
should be the ``target error'' for the characterization 
of the local geo-neutrino flux at 
each detector site. Much larger errors would hide information coming
from farther reservoirs, and in particular from the mantle.

\subsection{Provisional assumptions}

In order to make provisional numerical estimates, we make the following assumptions
for local HPE abundances in the crust: (1) we assume the same numerical values and errors
as for the average upper, middle, and lower crust estimates \cite{Rudn} discussed in Sec.~3.2,
except for the Kamioka site where the average upper crust abundances are taken from the
thorough geochemical study in \cite{Japa}; (2) we assume that the correlations between
local and global abundances, as well between local crust layers, are zero; 
(3) we assume a plausible (but arbitrary) 
hierarchy of correlations between HPE abundances in each layer:
$\rho(\mathrm{U,Th})=0.8$,  $\rho(\mathrm{U,K})=0.7$, $\rho(\mathrm{K,Th})=0.6$,
implying that Th is a good proxy for U and that K is a somewhat worse proxy
for both U and Th
(as it generally happens in other
reservoirs). Further comments about such choices are given elsewhere \cite{Ours}.
As previously remarked, the admittedly arbitrary assumptions characterizing local
contributions to geo-$\nu$ fluxes
can and should be improved by dedicated inter-disciplinary studies.

\section{Forward propagation of uncertainties}
\label{sec:forw}

We have described in Sec.~2 a possible path towards the definition of a GNSM, i.e.,
of a set of HPE 
abundances, errors and correlations in a given partition of the Earth into
global and local reservoirs. We now show examples of propagation of such uncertainties,
according to Eqs.~(\ref{PQ1})--(\ref{rhoPQ}).

\subsection{Errors and correlations at a specific site (Kamioka)}

Figure~3 shows our estimated geo-neutrino event rates from U and Th decays at the
Kamioka site (including neutrino oscillations with $\langle P_{ee}\rangle=0.57)$,
superposed to the same experimental (gaussian) contours as in Fig.~1. The numerical
values for the GNSM predictions are: 
\begin{equation}
R_\mathrm{U}=24.9\pm 2.0\mathrm{\ TNU},\ R_\mathrm{Th}=6.7\pm 0.5 \mathrm{\ TNU},\
\rho = +0.902\ .  
\end{equation} 
The correlation between the theoretical rates is positive, since Th and U are 
good proxies of one another in each reservoir. The total (U+Th) rate in Kamioka is also
positively correlated with the total (U+Th+K) radiogenic heat $H$ in the Earth. We estimate:
\begin{equation}
R_\mathrm{U+Th}=31.6\pm 2.5\mathrm{\ TNU},\ H=21.1\pm 3.0 \mathrm{\ TW},\
\rho = +0.858\ .  
\end{equation} 
The strong correlation between $R_\mathrm{U+Th}$ and $H$ implies that
a precise measurement of the former would yield a robust
constraint on the latter.
Unfortunately, the experimental errors in Fig.~3 are still much larger
than the theoretical (GNSM) ones, implying
that, at present, the first KamLAND data do not significantly constrain plausible Earth 
models and the associated radiogenic heat 
(see also \cite{Heat}). Patient accumulation of statistics, significant reduction
of background and systematics, and new independent experiments, are required
to test and constrain typical Earth model predictions.

\begin{figure*}
\centering
  \includegraphics[width=0.5\textwidth]{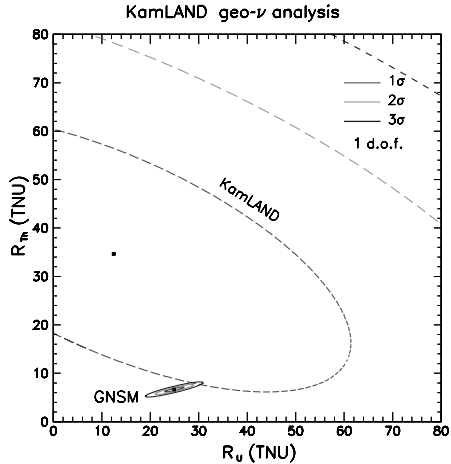}
\caption{U and Th geo-neutrino event 
rate predictions from our tentative GNSM at Kamioka (small ellipses with positive correlations),
superposed to the same KamLAND experimental constraints as in Fig.~1. In both cases, the
1, 2, and $3\sigma$ contours are shown. Current 
experimental errors appear to be significantly larger than the ``theoretical'' GNSM 
ones.}
\label{fig:3}       
\end{figure*}

\subsection{Errors and correlations among different sites}

Table~3 shows our estimates for the total (U+Th) rates (central values and
$\pm1\sigma$ error) at different possible detector sites, together with their
correlation matrix. Correlations are always positive (when one rate increases,
any other is typically expected to do the same), but can either be strong
(such as between Gran Sasso and Pyh\"asalmi, both located in somewhat similar CC
settings), or relatively weak (such as between any ``continental crustal site'' and 
the peculiar ``oceanic site'' at Hawaii, which sits on the mantle). Such correlations should be 
taken into account in the future, when data from two or more detectors
will be compared with Earth models.

\begin{table}[t]
\caption{\label{ForwardTot} Expected total neutrino event rates (U+Th), together with their errors and correlations, as calculated for
different sites within the GNSM, assuming $\langle P_{ee}\rangle=0.57$. }
\centering
\resizebox{\textwidth}{!}{
\begin{tabular}{lc|ccccccc}
\hline\noalign{\smallskip}
					& Rate (U+Th)  	& \multicolumn{6}{c}{Correlation matrix} \\
Site					&$\pm1\sigma$ (TNU)		& Kam.	& Gra.	& Sud.	& Haw.	& Pyh.	& Bak.	\\[1mm]	
\tableheadseprule\noalign{\smallskip}
Kamioka (Japan)		& $31.60\pm2.46$			& 1.000 	& 0.722	& 0.649	& 0.825	& 0.630	& 0.624 	\\
Gran Sasso (Italy)	& $40.55\pm2.86$			& 		& 1.000	& 0.707	& 0.641	& 0.734	& 0.700 	\\
Sudbury (Canada)		& $47.86\pm3.23$			& 		& 		& 1.000	& 0.554	& 0.688	& 0.652 	\\
Hawaii (USA)			& $13.39\pm2.21$			& 		& 		& 		& 1.000	& 0.484	& 0.510 	\\
Pyh\"asalmi (Finland)& $49.94\pm3.45$			& 		& 		& 		& 		& 1.000	& 0.692 	\\
Baksan (Russia)		& $50.73\pm3.41$			& 		& 		& 		& 		& 		& 1.000 	\\
\noalign{\smallskip}\hline
\end{tabular}
}
\end{table}

\section{Backward update of the GNSM error matrix}
\label{sec:back}

As shown in Fig.~3, the first KamLAND data do not yet constrain our tentative GNSM. 
However, it is tempting to investigate the impact of future, high-statistics
and multi-detector geo-neutrino data on the model. 
In particular, one might
try to estimate what are the HPE abundances which best fit {\em both\/}
the starting GNSM {\em and\/} a set of prospective, hypothetical experimental data
(``backward'' update of GNSM errors).
It can be shown that the covariance formalism allows
to reduce this problem to matrix algebra \cite{Ours}. 

Here we give just a relevant 
example of possible results, in an admittedly optimistic future scenario
where all 6 detectors in Table~3 are operative and collect separately
U and Th events for a total exposure of 20 kilo-ton year, at exactly 
the predicted rate, with no background and no systematics. In such
scenario, the mantle and BSE uranium abundance errors would be reduced
as in Fig.~4, which should be compared with the previous 
estimates in Fig.~2. It can be seen that, in principle, the depicted
scenario might allow to reject at $\gg 3\sigma$ the case $a_\mathrm{LM}=a_\mathrm{UM}$,
i.e., of global mantle convection, which would be a really relevant
result in geophysics and geochemistry. Needless to say, 
more realistic (and less optimistic) simulations of prospective data
need to be performed in order to check if similar goals can be experimentally
reached.
In any case, our approach may provide a useful template for such numerical studies.%
\begin{figure*}
\centering
  \includegraphics[width=.99\textwidth]{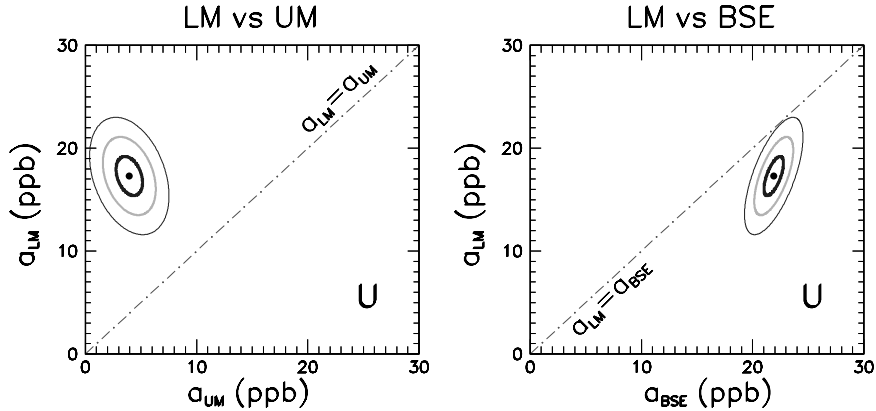}
\caption{As in Fig.~2, but including constraints from a hypothetical
data set from six detectors running for 20 kTy in ideal conditions. See
the text for details.}
\label{fig:4}       
\end{figure*}

\section{Conclusions and prospects for further work}
\label{sec:conc}

In this contribution to the Hawaii Workshop on Neutrino Geophysics (2005) we have
briefly presented a systematic approach to the ubiquitous issue of covariances
in geo-neutrino analyses. Correlations among the abundances of (U,Th,K) in each
reservoir and among different reservoirs, as well as covariances between
any two linear combinations of such abundances (including neutrino fluxes,
event rates, heat production rates) have been treated in a statistically
consistent way. A tentative Geo Neutrino Source Model (GNSM)---embedding 
a full error matrix for the (U,Th,K) abundances in relevant local and
global reservoirs---has been built, based on published data (when available) and on
supplementary assumptions (when needed). The construction of the GNSM highlights
some crucial issues that should be solved by dedicated studies, in order
to get the most from future geo-neutrino data.
Applications of our approach have been given in terms of predictions for future
experiments (``forward'' propagation of errors) and of GNSM error
reduction through prospective data (``backward'' update of uncertainties).
Inter-disciplinary studies of more refined geochemical and geophysical 
Earth models, and of future possible observations of geo-neutrino signals,
are needed to quantify more realistically both the assumed uncertainties and
the future impact of geo-neutrino
data in Earth sciences.


\begin{acknowledgements}
E.L.\ thanks the organizers of the Hawaii Workshop on Neutrino Geophysics (2005), where results
of this work were presented, for
kind hospitality and support. E.L.\ also thanks and W.F.\ McDonough and G.~Fiorentini
for interesting discussions about the BSE model. This work is supported in part by the Italian
Istituto Nazionale di Fisica Nucleare (INFN) and Ministero dell'Istruzione, Universit\`a 
e Ricerca (MIUR) through the ``Astroparticle Physics'' research project.
\end{acknowledgements}


\end{document}